\documentclass[12pt,a4]{article}
\usepackage{amsmath}
\usepackage{amssymb}
\usepackage{float}
\usepackage{amsfonts}
\usepackage{array}
\usepackage{graphicx}
\usepackage{cite}
\usepackage{mathrsfs}
\usepackage{multirow}
\usepackage{setspace}   
\usepackage{siunitx}
\usepackage[utf8]{inputenc}
\usepackage[colorlinks=true, linkcolor=blue, citecolor=blue, urlcolor=blue]{hyperref}
\usepackage{url}
\newcommand{\bea}{\begin{eqnarray*}}
\newcommand{\eea}{\end{eqnarray*}}
\newcommand{\beao}{\begin{eqnarray}}
\newcommand{\eeao}{\end{eqnarray}}

\begin{document}
\doublespacing 
\title{Bouncing Cosmology in 1+1 Dimensions}

\author{Hagar Ariela Meir\\
\small Department of Physics, Ben-Gurion University, Beer-Sheva 8410501, Israel}
\maketitle

\begin{abstract}
 In this paper, I construct a bouncing cosmology by considering the backreaction of the winding condensate on a 1+1 dimensional cosmological model with a periodic spatial coordinate. I based my work on previous results that considered the backreaction of the winding condensate on a 1+1 dimensional Euclidean black hole. This cosmological model is obtained as an analytic continuation of a Euclidean black hole. I solved the equations and obtained non-singular solutions at near-Hagedorn temperatures, both numerically and analytically. To remain within the weak coupling regime, it is necessary to connect two solutions; otherwise, the dilaton, which determines the string coupling, would grow quadratically. This connection is achieved through a smooth coordinate transformation, ensuring the model's validity. As a result, the model becomes geodesically complete and non-singular. The connection is made at a time in which the curvature is small, thereby avoiding higher-order $\alpha'$ corrections.
 
\end{abstract}

\section{Introduction}
Recent studies of non-singular Black Holes (BHs) have used the Euclidean BH, dilaton, and winding condensate framework \cite{Itzhaki:2018rld, giveon2015stringy, Giveon_2014, Giveon_2014_2, Giveon_2016, Ben_Israel_2015, Ben_Israel_2017, Martinec_2019}, focusing on BHs at the near horizon limit, beyond the horizon, and in the near Hagedorn temperature approximation. The Penrose-Hawking energy condition violation and the possibility of a non-singular universe have also been discussed \cite{Itzhaki_2021}.

Building on \cite{Puncture}, which employed an action involving the string condensate wrapped around the cyclic time coordinate in a two-dimensional Euclidean BH, I used the string's action with dilaton gravity and a 1+1 FRW metric. This approach resulted in four equations of motion (EOM) that were solved numerically.

 I constructed a cosmological model analogous to the punctured Euclidean BH in \cite{Puncture}, which was found to be non-singular. By performing an analytic continuation of space and time, to obtain an FRW cosmological model, I ensured the action retains its form, yielding solutions mirroring those of the punctured BH. This process led to a non-singular expanding model of the universe.

 As the curvature grows, higher-order contributions to the effective action become significant. When the curvature or the gradients of the background fields are comparable to the fundamental string length, the perturbative expansion fails. An exact conformal field-theory model, which includes all order corrections, should be used instead \cite{gasperini_2007_chapter2_3}.

To ensure this model's validity, higher-order corrections must be subdominant, setting a limit on the dilaton field. Due to this requirement, I connected two solutions at a point where the dilaton reaches its maximal value. This creates a bouncing cosmological model, connecting the expanding model with a time-reflected contracting solution. A coordinate transformation of time smoothly linked the solutions at the origin, yielding a bouncing cosmology.

The ``Big Bounce" concept explores pre-Big Bang conditions and addresses singularities in standard cosmology through quantum mechanics \cite{Conzinu:2023fth,Veneziano:2003sz,NOVELLO_2008}. General relativity (GR) inadequately describes the universe's earliest moments near the Big Bang due to singularities \cite{Robson19}. The Big Bounce offers an alternative by transitioning from a contracting to an expanding universe, avoiding singularities.

Research on non-singular universes includes models of accelerated collapse to expansion and cyclic universes \cite{NOVELLO_2008, Brandenberger_2017}.

Different modified theories of gravity have been proposed \cite{CANTATA:2021ktz}, including models such as Brans-Dicke gravity  \cite{Brans:1961sx}, modified $f(T)$ teleparallel gravity \cite{Cai:2017avs}, Kaluza-Klein theories
\cite{Castillo-Felisola:2016kpe,Shaikh:2022meo,Overduin:1997sri}
, and Hořava-Lifshitz gravity \cite{Sotiriou:2010wn}.
Another suggestion is theories that preserve the core principles of Einstein's General Relativity, with modified $f(R)$ gravity \cite{Sotiriou:2008rp}, where the curvature is replaced with a scalar function of it, $f(R)$.

Higher-order gravitational theories, such as $R^2$ and $R_{\mu\nu}R^{\mu\nu}$, can violate energy conditions, enabling a bounce \cite{DAVIES1977108, Damour_1994, PhysRev.125.1727,Singh_2023,Pavlovic:2017umo}. Scalar field theories also support bouncing cosmologies through energy condition violations \cite{PhysRevD.12.3077, Bayin_1994, Gasperini_2003}.

Modifications to Maxwell's electrodynamics in the FRW model can produce non-singular solutions in classical GR \cite{PhysRevD.20.377, PhysRevD.65.063501}. String theory, with its higher-dimensional models, alters the Friedmann equation on a 3-brane, potentially leading to a bounce \cite{Germani_2008}. Treating the cosmological constant $\Lambda$ as time-varying can introduce a dynamic Friedmann equation, suggesting a bouncing cosmology \cite{KIRZHNITS1976195, ADLinde_1979, Alvarez:2023gfn}.

The classical Big Bang Theory predicts a singularity where Einstein equations cease to be applicable. In contrast, a bouncing model avoids this singularity, allowing geodesics to trace back through time before the bounce, providing a compelling and innovative alternative to the Big Bang model \cite{NOVELLO_2008, Brandenberger_2017}.

In the construction of this model, the non-singular solution was obtained without including higher orders of curvature, or giving the cosmological constant a time-changing property. 
The result was given merely from a violation of the general covariance principle.
I utilized equations found from an action that involved a non-local term (the Horowitz-Polchinsky term), which caused a violation of the general covariance principle, an essential assumption in the Hawking-Penrose singularity theorem. This violation affects the theorem's prediction of singularity.

\section{Cosmological Puncture}

The HP action (Eq.(\ref{eq:H.P action})) is an action that describes a single highly excited string by gradually increasing the string coupling from weakly coupled matter, at a critical compactification radius.
In this action, a scalar with winding number one becomes tachyonic (the mass term squared becomes negative) for $\beta<\beta_H$.

\begin{equation}\label{eq:H.P action}
    S = \int d^d x \sqrt{g} e^{-2\Phi} \left[g^{\mu\nu} \partial_{\mu}\chi \partial_{\nu} \chi^{*} + \frac{\beta^2g_{\tau\tau} - \beta_{H}^2}{4\pi^2\alpha'^2}\chi\chi^* - \frac{1}{2\kappa^2}\left(R - 2\Lambda + 4g^{\mu\nu}\partial_{\mu}\Phi \partial{\nu}\Phi\right)\right].
\end{equation}

where $\kappa^2 = 8\pi G_N$, $\beta =2\pi\sqrt{\alpha' k}$, $\beta_H = 2\pi\sqrt{\alpha' 2}$, and $\Lambda = -\frac{2}{\alpha' k}$. This action is constructed from the dilaton-graviton action and the winding condensate action.

The dilaton field is denoted by \(\Phi\), which is the scalar field that determines the string coupling, \(g_s = e^{\Phi}\). 

The fields \(\chi\) and \(\chi^*\) represent the \(\pm 1\) winding condensates of a string wrapped around Euclidean time. 

The metric tensor is denoted by \(g_{\mu \nu}\), with \(g_{\tau\tau}\) being its \(\tau - \tau \) component, where $\tau$ is the periodic time coordinate.

The parameter \(\beta\) denotes the periodicity of the \(\tau\) circle, such that \(\tau \sim \tau + \beta\). At finite temperature in thermal equilibrium, \(\beta\) is identified with the inverse Hawking temperature of the black hole, expressed as \(\beta = \frac{1}{T} = 4 \pi R_s\), where \(R_s\) is the Schwarzschild radius.

The constant \(k\) determines the scale of the cigar and corresponds to the level of the \(SL(2, \mathbb{R}) \) current algebra \cite{giveon2015stringy}.

The critical behavior as $\beta \rightarrow \beta_H$ is described by an effective field theory of a single complex scalar field in two space-time dimension \cite{Kogan:1987jd,ATICK1988291}.

This arises from a Hagedorn singularity at a critical compactification radius, where a scalar with winding number one becomes tachyonic (the mass term squared becomes negative) for $\beta<\beta_H$.
The field $\chi$ has a winding number one, while $\chi^*$ has a winding number minus one.

In \cite{Puncture}, they used the HP action and the corresponding EOM and found a nonsingular solution to a Euclidean BH, a puncture.
To find a cosmological solution that resembles the puncture, I performed an analytic continuation to the Euclidean BH parameters, the corresponding action of self-gravitating fundamental strings constructed by Horowitz and Polchinski \cite{Horowitz_1998}, and the equations of motion(EOM) found in \cite{Puncture}.

\subsection{The Puncture in the Euclidean Black Hole}

The HP term in the action in Eq.(\ref{eq:H.P action});  $\dfrac{\beta ^2 g_{\tau \tau}}{4\pi{\alpha'} ^2}\chi\chi^*$ is not covariant or not local.

The appearance of this term in the action results in its appearance in the EOM obtained by varying the action in Eq.(\ref{eq:H.P action}) and in the stress-energy tensor derived from the Lagrangian.

This violates the assumption of Penrose and Hawking regarding the Einstein equations in \cite{Hawking:1970zqf}, which allows the result of a non-singular solution.

In \cite{Puncture}, this action was used with the metric of a Euclidean BH in Eq.(\ref{eq:Euclidean B.H}) to obtain the equations of motion.
\begin{equation}\label{eq:Euclidean B.H}
    ds^2 = d\rho^2 +h^2(\rho)d\tau^2
\end{equation}

As a result, the relevant equations of motion are;

\begin{equation}\label{eq:eom of euclidean bh}
\begin{aligned}
\sqrt{k} h' &= \sqrt{k} h \phi' + 1 \\
\phi' &= -\sqrt{k} h  \left( F + \frac{1}{k} \right) \\
F' &= -2 \sqrt{k} h F ,
\end{aligned}
\end{equation}

where $F=\chi\chi^*$, which is the field that describes instant folded strings as explained in
\cite{StringyInformation} and \cite{Giveon:2020xxh}.
In \cite{Puncture}, the perturbative solutions for Eq.(\ref{eq:eom of euclidean bh}) yielded a puncture in the Euclidean BH.

\subsection{Friedmann Robertson Walker cosmology from the Euclidean BH}

I aimed to find a non - singular cosmology described by the Friedmann Robertson Walker(FRW) model.
To find such a model that would be non-singular, I performed an analytic continuation on the Euclidean BH metric given by Eq.(\ref{eq:Euclidean B.H}), defined by Eq.(\ref{eq:analytical continuation});

\begin{equation}\label{eq:analytical continuation}
    \left\{
    \begin{array}{ll}
    \tau \rightarrow i r \\
    \rho \rightarrow i t \\
    \Lambda \rightarrow -\Lambda \\
    h(\rho) \rightarrow i a(t) \\
    k \rightarrow -k \\
    \alpha' \rightarrow -\alpha' .
    \end{array}
    \right.
\end{equation}

This analytic continuation converts the 1+1 dimensional Euclidean BH metric into the 1+1-dimensional FRW metric in Eq.(\ref{eq:FRW metric}), providing us with a cosmology metric:

\begin{equation}\label{eq:FRW metric}
    ds^2 = -dt^2 +a(t)^2dr^2.
\end{equation}

The roles of periodic time and space have been interchanged in this analytic continuation, implying that the strings are now wrapped around a periodic spatial coordinate with periodicity of $\beta = 2\pi\sqrt{k}$.


Considering the analytic continuation of $\tau$ and $\rho$, any first-order derivative of the fields $\chi$ and $\Phi$ is multiplied by $-i$. Consequently,
The Ricci scalar $R$ is calculated as:

\begin{equation}\label{eq:Ricci scalar Euclidean BH}
    R=-2\frac{h''(\rho)}{h(\rho)} \rightarrow 2\frac{\ddot{a(t)}}{a(t)} .
\end{equation}

By implementing all these transformations, the eom keep the form of Eq.(\ref{eq:eom of euclidean bh}):

\begin{equation}\label{eq:a(t) simplified}
    \dot{a}(t) = a(t)\dot{\Phi}(t) + \frac{1}{\sqrt{k}},
\end{equation}
\begin{equation}\label{eq:phi(t) simplified}
    \dot{\Phi}(t) = -a(t) \left( \sqrt{k} F(t) +\frac{1}{\sqrt{k}}\right),
\end{equation}
\begin{equation}\label{eq:chi(t) simplified}.
    \dot{F}(t) = -2\sqrt{k}a(t)F(t) .
\end{equation}

 I used the following asymptotic approximations, in the region $1\ll t \ll \sqrt{k}$ , as boundary conditions for $a(t)$, $\Phi(t)$ and $F(t)$, to numerically solve Eq.(\ref{eq:a(t) simplified})-eq:(\ref{eq:chi(t) simplified}):

\begin{equation}\label{eq:a 1<<t<<sqrtk}
    \sqrt{k } a(t) = t ,
\end{equation}

\begin{equation}\label{eq:phi(t) 1<<t<<sqrtk}
    \Phi(t) = \Phi_0 ,
\end{equation}

\begin{equation}\label{eq:chi(t) 1<<t<<sqrtk}
  F(t) = A^2 e^{-t^2} .
\end{equation}

\begin{figure}[H]
  \centering
  \includegraphics[scale=0.6]{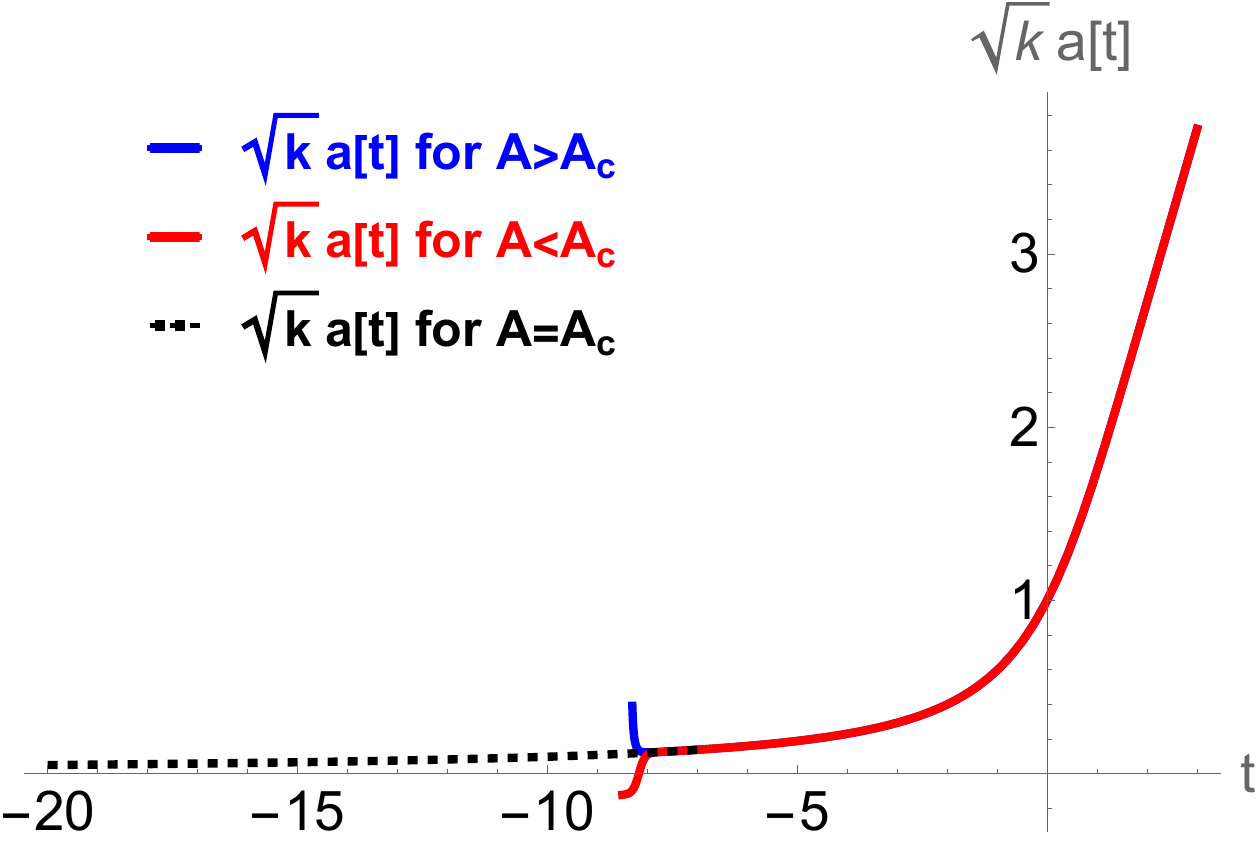}
  \caption{$\sqrt{k}a(t)$ as a solution for Eqs.(\ref{eq:a(t) simplified})-(\ref{eq:chi(t) simplified}) with the boundary conditions in Eq.(\ref{eq:a 1<<t<<sqrtk})-(\ref{eq:chi(t) 1<<t<<sqrtk}) }
  \label{fig:a(t) solution}
\end{figure}

In Fig.\ref{fig:a(t) solution}, the solution of $\sqrt{k}a(t)$ is obtained numerically by solving the EOM for a large $k$ with the boundary conditions given by Eqs.(\ref{eq:a 1<<t<<sqrtk})-(\ref{eq:chi(t) 1<<t<<sqrtk}). The solution is shown for amplitudes $A$ of $\chi(t)$, both above (blue curve) and below (red curve) the critical amplitude found in \cite{Puncture}, $A_c = e^{-\gamma/2}$, where $\gamma = 0.57721...$ is Euler's constant. As expected, the puncture solution $\sqrt{k}a(t)=-\frac{1}{\rho-1}$ smoothly continues to the solution.

Moreover, to validate the accuracy of the fittings in Eqs.(\ref{eq:a(t) t<<-1})-(\ref{eq:chi(t) t<<-1}) to the numerical solutions, I performed a numerical calculation, with boundary conditions set in $t\ll -1$, using the asymptotic solutions as boundary conditions.
\begin{equation}\label{eq:a(t) t<<-1}
    \sqrt{k}a(t) = -\frac{1}{t-1} ,
\end{equation}
\begin{equation}\label{eq:phi'(t) t<<1}
    \Phi'(t) = t-1 ,
\end{equation}
\begin{equation}\label{eq:chi(t) t<<-1}
    F(t) = (-t+1)^2 .
\end{equation}

The numerical calculations yielded results consistent with the asymptotic solutions of Eqs.(\ref{eq:a(t) t<<-1})-(\ref{eq:chi(t) t<<-1}) in the region of $t\ll-1$ and the asymptotic solution of Eqs.(\ref{eq:a 1<<t<<sqrtk})-(\ref{eq:chi(t) 1<<t<<sqrtk}) in the region where $t \gg 1$, as expected.
In the following figures, it can be seen how the solutions perfectly match the solutions found using the boundary condition Eqs.(\ref{eq:a 1<<t<<sqrtk})-(\ref{eq:chi(t) 1<<t<<sqrtk}).


\begin{figure}[H]
  \centering
  \includegraphics[scale=0.7]{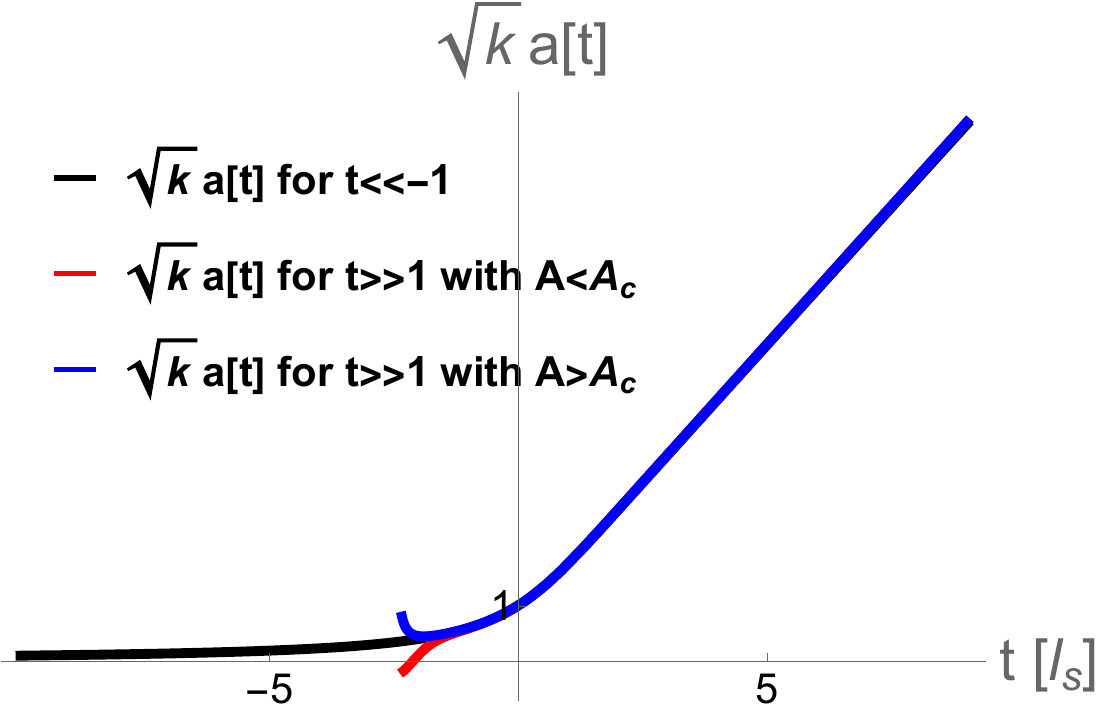}
  \caption{A plot of $\sqrt{k}a(t)$ as a numerical solution for Eqs.(\ref{eq:a(t) simplified})-(\ref{eq:chi(t) simplified}) with the boundary conditions in Eqs.(\ref{eq:a(t) t<<-1})-(\ref{eq:chi(t) t<<-1}) in the blue line, and with the boundary conditions of Eqs.(\ref{eq:a 1<<t<<sqrtk})-(\ref{eq:chi(t) 1<<t<<sqrtk}) in the blue and red lines. }
  \label{fig:a(t)}
\end{figure}

\begin{figure}[H]
  \centering
  \includegraphics[scale=0.7]{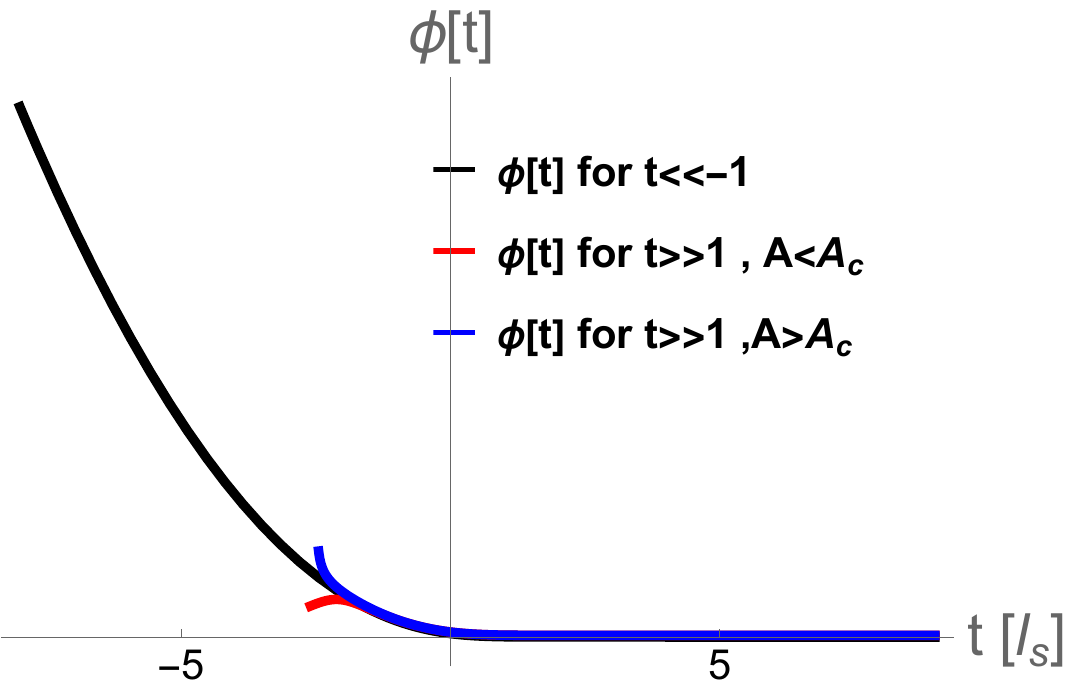}
  \caption{A plot of $\Phi(t)$ as a numerical solution for Eqs.(\ref{eq:a(t) simplified})-(\ref{eq:chi(t) simplified}) with the boundary conditions in Eq.(\ref{eq:a(t) t<<-1})-(\ref{eq:chi(t) t<<-1}) in the blue line, and with the boundary conditions of Eqs.(\ref{eq:a 1<<t<<sqrtk})-(\ref{eq:chi(t) 1<<t<<sqrtk}) in the blue and red lines.}
  \label{fig:phi(t)}
\end{figure}

\begin{figure}[H]
  \centering
  \includegraphics[scale=0.7]{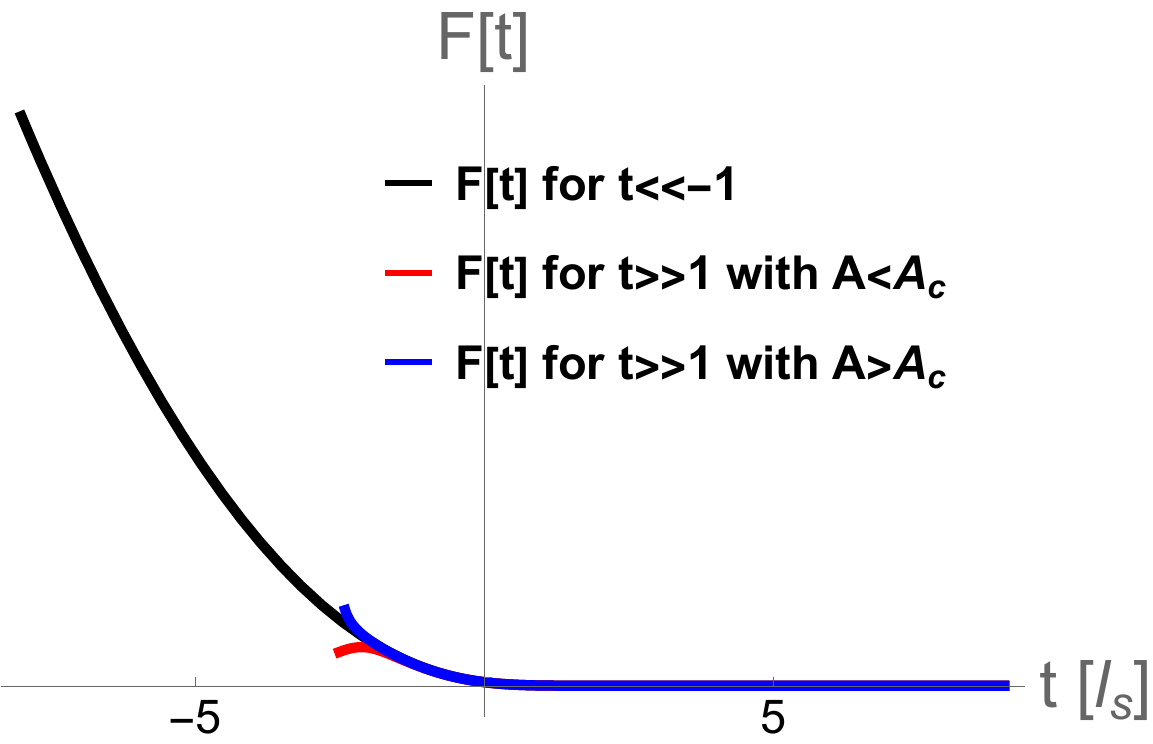}
  \caption{A plot of $F(t)$ as a numerical solution for Eqs.(\ref{eq:a(t) simplified})-(\ref{eq:chi(t) simplified}) with the boundary conditions in Eqs.(\ref{eq:a(t) t<<-1})-(\ref{eq:chi(t) t<<-1}) in the blue line, and with the boundary conditions of Eqs.(\ref{eq:a 1<<t<<sqrtk})-(\ref{eq:chi(t) 1<<t<<sqrtk}) in the blue and red lines.}
  \label{fig:chi(t)}
\end{figure}

The transformation of time; $t\rightarrow -t$ does not change the eom, so that  $a(-t)$, $\Phi(-t)$, and $F(-t)$ are also solutions. By using the time-reflected boundary conditions (for $-t$ instead of $t$), the time-reflected solutions can be obtained.

\section{Connection at the beginning(end) of time}

My solutions for the FRW cosmological model in Eq.(\ref{eq:FRW metric})  establish a non-singular cosmological model, as the line element is defined everywhere and the curvature is smooth and finite.

However, the HP action is only valid for small values of the string coupling. Since the string coupling is determined by the exponential of the dilaton field, \(\Phi(t)\) must approach a finite value.

The asymptotic solution; $\Phi'(t)$ in Eq.(\ref{eq:phi'(t) t<<1}), as well as in Fig.\ref{fig:phi(t)} imply that $\Phi(t)$ grows quadratically as $t \rightarrow -\infty$. 

This means that this solution by itself can not be a valid cosmological model, therefore we must construct a model such that the dilaton field reaches a finite value.

The solutions I found (Fig. \ref{fig:a(t)} -  \ref{fig:chi(t)})  their reflection ( $-t$ ) are equally valid solutions. Therefore, I aimed to connect them at a certain point in time $t_1$.

\subsection{Coordinate transformation $f(v) = t$}\label{sec:coordinate transformation}
Simply connecting $a(t)$, $\Phi(t)$, and $F(t)$ at a point $t_1$ would make derivatives discontinuous. To overcome this, I implemented a coordinate transformation $t=f(v)$ such that $f(0)=t_1$, $f'(0)=0$, and $f''(0) =0$. Furthermore, I choose $f$ such that above an arbitrarily chosen $v_0$ and below $-v_0$, the metric is exactly a 2-dimensional FRW model.

The transformation is defined as follows:

\begin{equation}\label{eq:f(v) arbitrary v0 and t_1}
     f(v)=\left\{
    \begin{array}{ll}
        -v + t_1 & v\leq -v_0 \\
       t_1 +\frac{6}{v_0^2}|v|^3 -\frac{8}{v_0^3}v^4+\frac{3}{v_0^4}|v|^5  &   -v_0\leq v \leq v0 \\
        v +t_1 & v_0 \leq v  
    \end{array}
    \right.   .
\end{equation}

The modified line element after the transformation:

\begin{equation}\label{eq:metric after coordinate transformation}
    ds^2 = -f'(v)^2dv^2 +a(f(v))^2 dr^2.
\end{equation}

As seen from Eq.(\ref{eq:f(v) arbitrary v0 and t_1}), the line element coincides with Eq.(\ref{eq:FRW metric}) above $v_0$ and below $-v_0$.

\begin{figure}[H]
  \centering
  \includegraphics[scale=0.6]{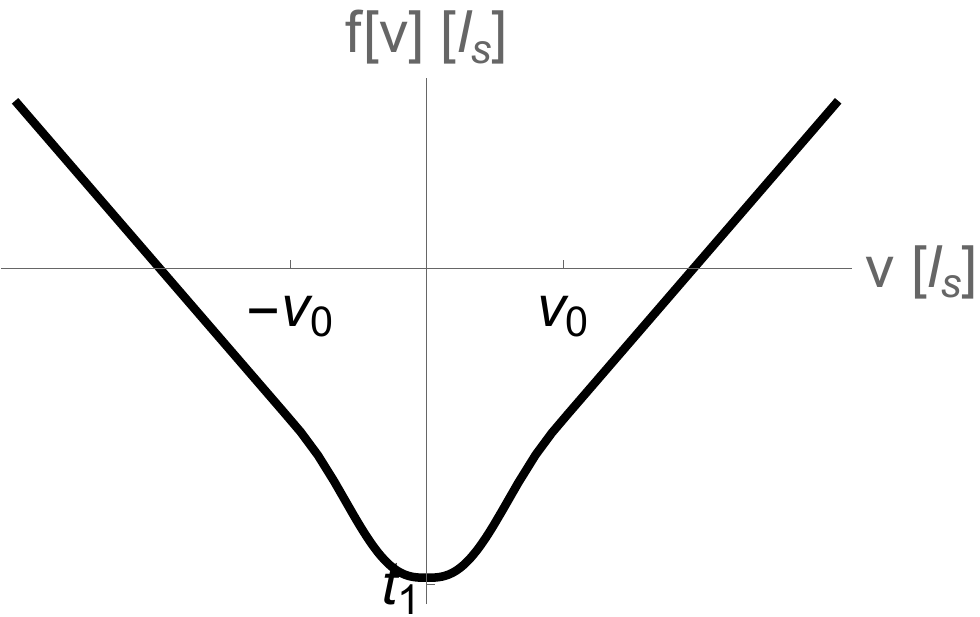}
  \caption{$f(v)$, the coordinate transformation in Eq.(\ref{eq:f(v) arbitrary v0 and t_1}) for $t_1=-10$ and $v_0 = 5$.}
  \label{fig:f(v)}
\end{figure}

As an example, I chose the specific values: $t_1 = -10$ and $v_0 = 5$.
The coordinate transformation $f(v)$ is illustrated in Fig.\ref{fig:f(v)}.

Upon implementing Eq.(\ref{eq:metric after coordinate transformation}), the solutions, $\Phi(f(v))$, $F(f(v))$, and $\sqrt{k}a(f(v))$ become continuous, and their first and second-order derivatives are continuous as well.

\begin{figure}[H]
  \centering
  \includegraphics[scale=0.6]{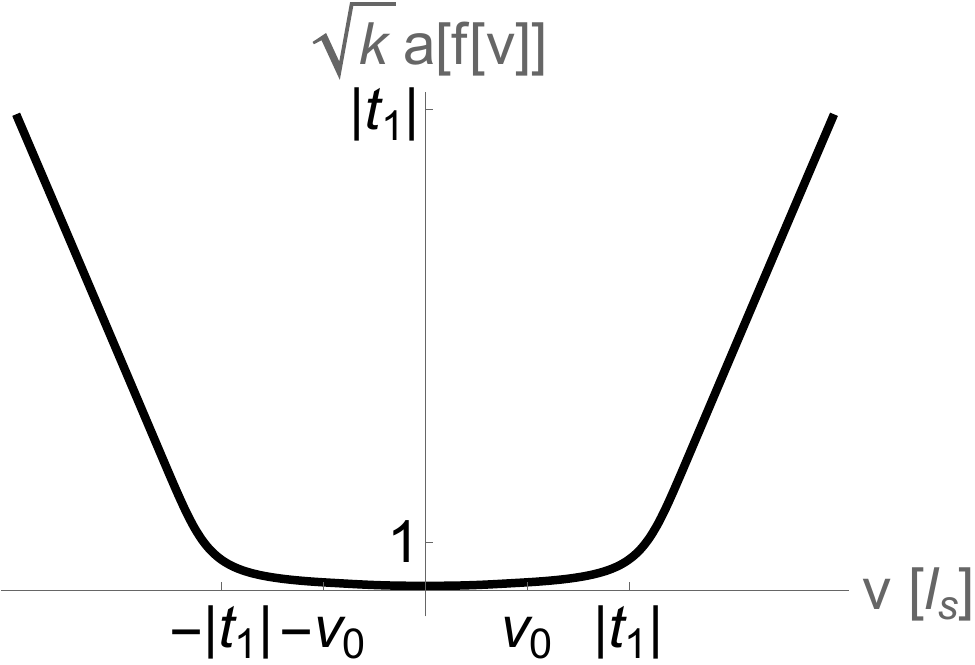}
  \caption{A plot of $a(f(v))$, as a numerical solution for Eqs.(\ref{eq:a(t) simplified})-(\ref{eq:chi(t) simplified}), and as function of $v$, with $t_1 = -10$ and $v_0 = 5$.}
  \label{fig:a(f(v))}
\end{figure}

\begin{figure}[H]
  \centering
  \includegraphics[scale=0.6]{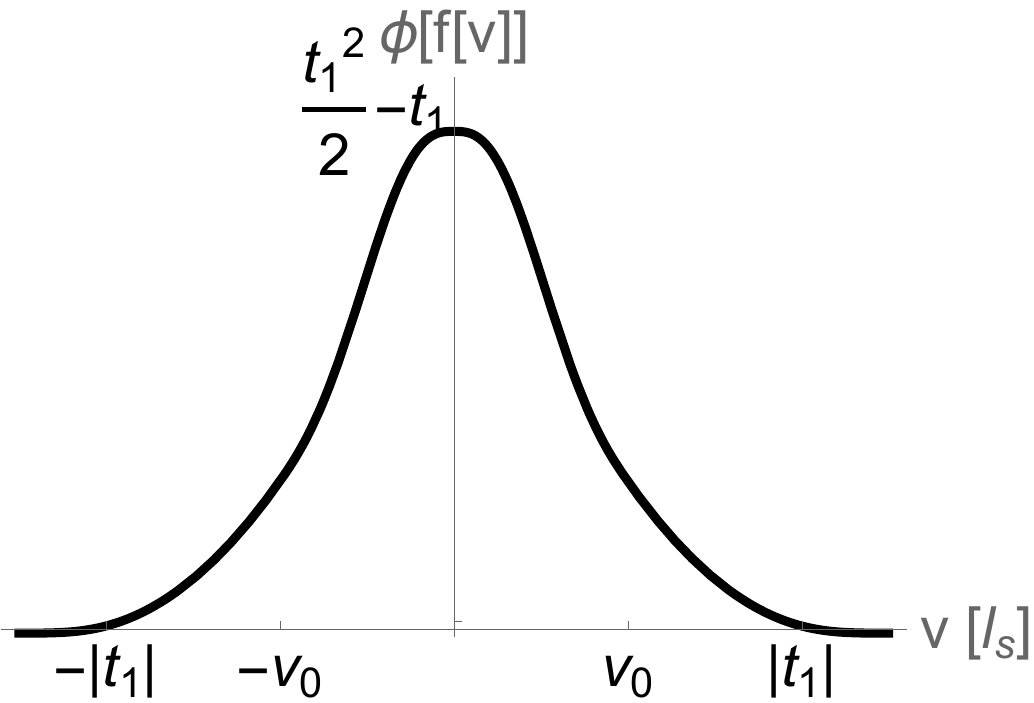}
  \caption{A plot of $\Phi(f(v))$, as a numerical solution for Eqs.(\ref{eq:a(t) simplified})-(\ref{eq:chi(t) simplified}), and as function of $v$, with $t_1 = -10$ and $v_0 = 5$.}
  \label{fig:phi(f(v))}
\end{figure}

\begin{figure}[H]
  \centering
  \includegraphics[scale=0.6]{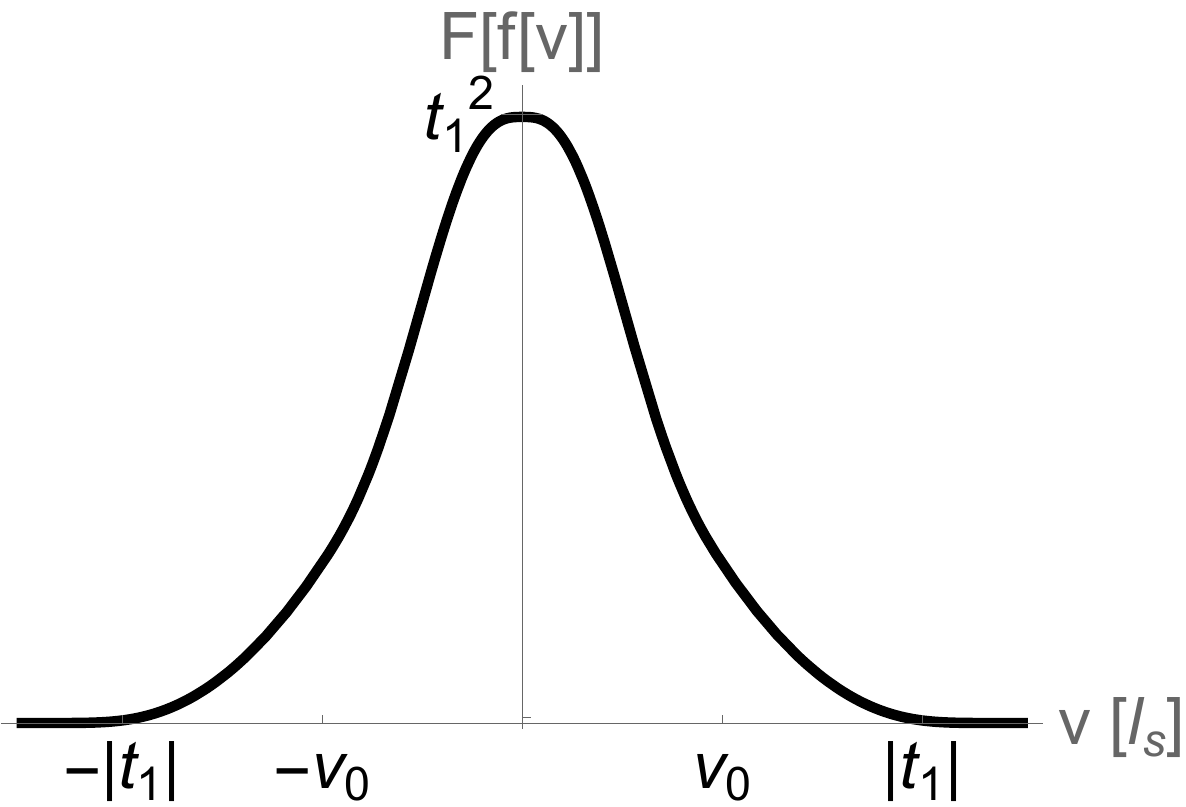}
  \caption{$F(f(v))$ as a numerical solution for Eqs.(\ref{eq:a(t) simplified})-(\ref{eq:chi(t) simplified}), and as function of $v$, with $t_1 = -10$ and $v_0 = 5$.}
  \label{fig:chi(f(v))}
\end{figure}

Choosing any $t_1<\infty$ and $\Phi_0$ results in a bouncing cosmology. However, for the theory to be valid everywhere in the leading order in $\alpha'$ and in the small coupling region, the choice of $t_1$ and $\Phi_0$ is significant.

Fig.\ref{fig:a(f(v))} and Fig.\ref{fig:chi(f(v))} demonstrate that $F$ reaches its maximal value when $a$ achieves its minimal value, which is consistent with what we would expect from having the mass term in the action. 
As $\beta^2 a^2 = 4\pi^2 k a^2$ becomes smaller than the inverse Hagedorn temperature squared $\beta_H^2 = 8\pi^2$, the $F$ field becomes tachyonic, leading to an increase in the folded string condensate, an expected behavior when crossing the Hagedorn temperature \cite{Giddings:1989xe}.

\subsection{Ricci curvature}

The expression for the Ricci curvature of the 1+1 FRW metric in Eq.(\ref{eq:FRW metric}) is given by:
\begin{equation}\label{eq:Ricci of t}
    R(t) = 2\frac{\ddot{a}(t)}{a(t)} .
\end{equation}
Since it is independent of coordinate transformation, we can express it in terms of our new time coordinate $v$ as follows:
\begin{equation}\label{eq:Ricci of v}
    R(v)=2\frac{\ddot{a}(f(v))}{a(f(v))}.
\end{equation}

For the point of connection to be at a location where the solution is valid according to first-order in $\alpha'$ perturbation theory, we need to choose the point of connection such that $R$ has a small value in units of $\frac{1}{l_s^{2}}$.

\begin{figure}[H]
  \centering
  \includegraphics[scale=0.65]{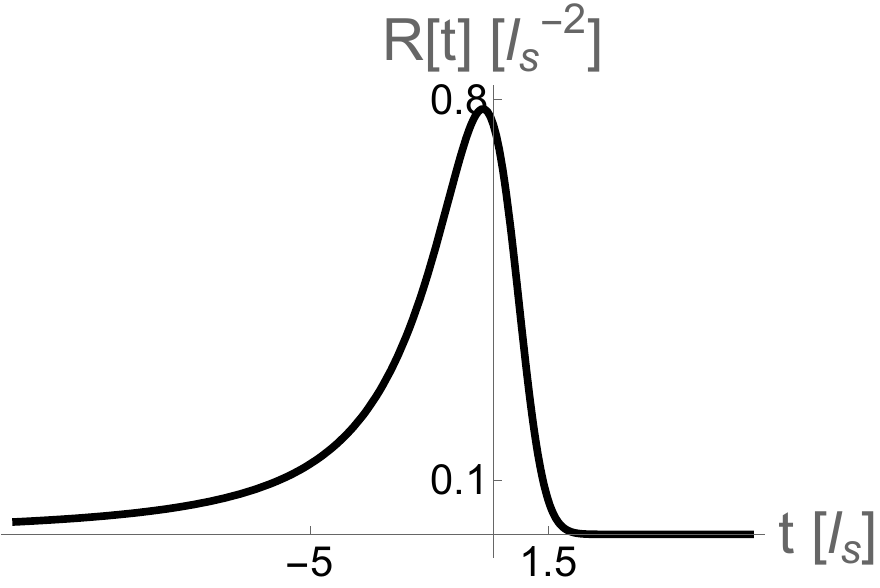}
  \caption{$R(t)$ The Ricci curvature as a function of $t$, found from Eq.(\ref{eq:Ricci of t}), and the numerical solution for $a(t)$.}
  \label{fig:R(t)}
\end{figure}

Fig.\ref{fig:R(t)} displays the Ricci scalar curvature as a function of $t$ for the original solution $a(t)$, as found in Fig.\ref{fig:a(t) solution}. The highest value of the curvature is approximately $R\sim 0.8$, and it is found near the origin, where the winding condensate grows and starts to influence the solutions.

The curvature scalar has this relatively high value only in a small region around the origin. I have solved the equations for $t\ll -1$ as well as for $t \gg 1$, and the two solutions are valid and smoothly connected. Therefore, I inferred that the choice of $t_1$ where $t\ll -1$ gives a valid solution to the leading order in $\alpha'$. Even though $R$ reaches a peak value of approximately $\sim 1$, it is still smaller than 1. The continuity of the connection between the solutions $t \ll -1$ and $t \gg 1$, and their validity, suggest that $\alpha'$ corrections would not significantly alter the behavior of the solution.

\subsection{Hubble parameter}
The Hubble parameter for an FRW metric is defined as $H=\frac{\dot{a}}{a}$.
I am redefining the Hubble parameter in this section to be 
\begin{equation}\label{eq:redefined Hubble}
    H(v) = \frac{1}{a} \frac{da}{dv}.
\end{equation}
Even though it does not represent the Hubble parameter, since this parameter is defined for a derivative in $t=f(v)$, while I defined it for $v$, from above the point $v_0$ and below $-v_0$, it is indeed the Hubble parameter of an FRW metric.

\begin{figure}[H]
  \centering
  \includegraphics[scale=0.6]{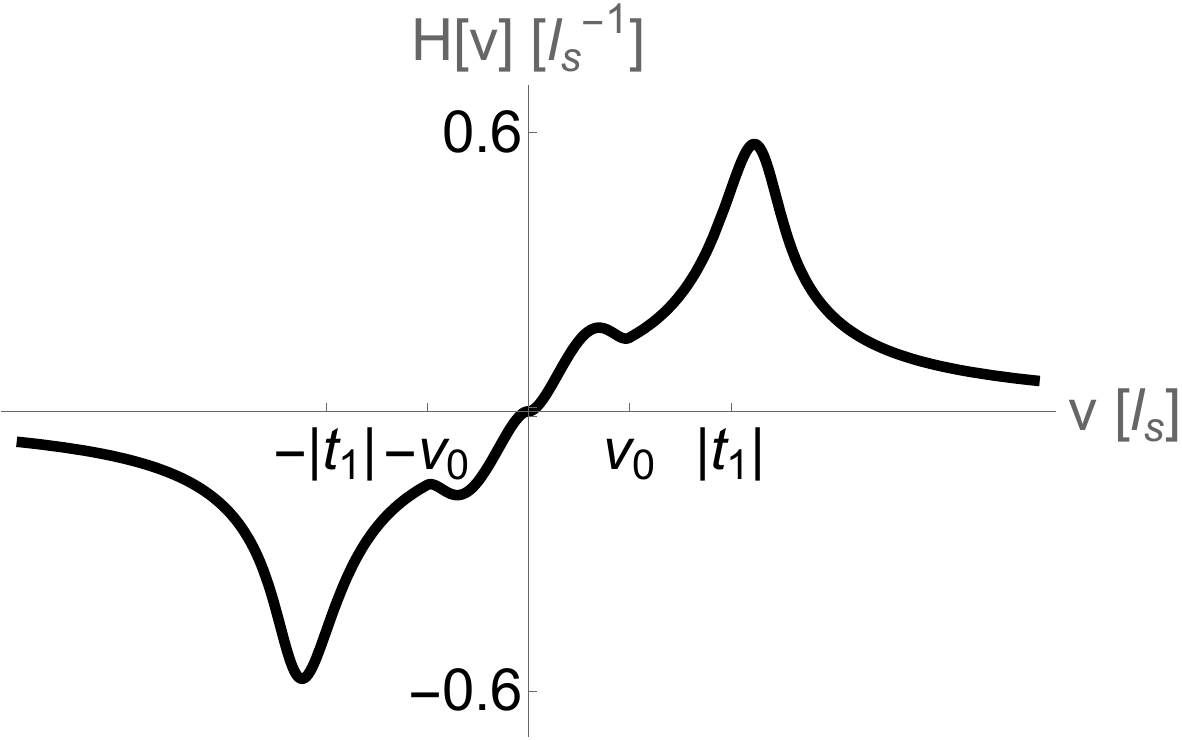}
  \caption{$H(v)$, The ``Hubble parameter'', as defined in Eq.(\ref{eq:redefined Hubble}).}
  \label{fig:H(v)}
\end{figure}

In Fig.\ref{fig:H(v)}, it can be seen that the Hubble parameter is negative for $v\ll-1$, as well as for $t\ll-1$ and that it is positive for $v\gg1$ as well as for $t\gg1$. This means that the average derivative of $H$ is positive:
\begin{equation}
    \bar{\dot{H}} \equiv \frac{\Delta H}{\Delta t} = \frac{\Delta H}{\Delta v} > 0.
\end{equation}

The Hawking-Penrose singularity theorem applies to GR and its assumptions.
I have already shown that the assumption of covariance and locality does not hold, and therefore singularity can be evaded.
As a result of this violation, I obtained a violation of the second assumption of the Hawking-Penrose singularity theorem, i.e. the 'energy condition', which for a flat FRW model states that $\dot{H} \leq 0$.
This means that as expected, from the violation of the general covariance principle, the energy condition of the Hawking-Penrose singularity theorem does not apply.

\section{Conclusion}

In this study, I successfully constructed a model that describes a bouncing universe in the coordinate $v$, achieving the intended goal.

I constructed a bounce in the coordinate $v$ and not in the original $t$ coordinate, through the coordinate transformation of $t=f(v)$ that I presented in Eq.(\ref{eq:f(v) arbitrary v0 and t_1}). The coordinates; $t$ and $v$ coincide from above and below at arbitrarily chosen points.
Also, all physical values discussed in this study are smooth and continuous; the curvature, the line element, the dilaton field, the winding condensate, and their derivatives.

Despite the limitations in our EOM due to the value of $R \approx 0.8$, it is important to note that this limitation only applies to a small range. As discussed earlier, the solutions smoothly connect around this range, maintaining a seamless transition. Consequently, including higher-order terms to complete this theoretical framework is not expected to significantly change our observed results.

Additionally, the perturbation method requires that the parameter $g_s^2$ be much smaller than 1. By selecting a very small value for $\Phi_0$, I established an upper limit for $g_s^2$ = $e^{{t_1}^2}$, ensuring it remains small for small $t_1$ \cite{gasperini2007dilaton}.

Furthermore, my proposed model matches the FRW framework from a time larger than $t=|t_1|$ and smaller than $t=-|t_1|$, the time $t_1$ can be chosen arbitrarily. By doing so, the model in the string frame matches the Einstein frame, in the region of $\Phi \sim \Phi_0$.
In these different scenarios, the region $-\infty < v \leq 0$ exhibits a contracting scale factor, while the $0 \leq v < \infty$ region displays an expanding scale factor, indicating that a positive slope for $H$ is maintained on average.

The freedom of choice for $t_1$ and $\Phi_0$, allows one to select values so that the near Hagedorn approximation Polchinski and Horowitz made (\cite{Horowitz_1998}) is valid, as well as our equations, which involve only leading order terms of $\alpha'$.

To summarize, this paper presents a consistent model of a bouncing 1+1 universe while addressing challenges related to the validity of the equations.

\section*{Acknowledgement}
I would like to thank Nissan Itzhaki for his innovative idea of employing previous results of the two-dimensional black hole to create a non-singular bouncing cosmology in 2 dimensions. Without his insight, this research would not have been possible. I also wish to acknowledge Yoav Zigdon for his valuable assistance with prior calculations that significantly contributed to this work. Lastly, I would like to thank my advisor, Ramy Brustein, for his consistent guidance and support throughout this research.
The research is supported by the German Research Foundation through a German-Israeli Project Cooperation (DIP) grant ``Holography and the Swampland'' and by VATAT (Israel planning and budgeting committee) grant for supporting theoretical high energy physics

\bibliographystyle{unsrt}

\end{document}